\documentclass[12pt,preprint]{aastex}

\slugcomment{Accepted for publication in The Astronomical Journal}

\shorttitle{\HI~in B2352+495}
\shortauthors{Araya et al.}

\def \Mo {$\,$M$_{\odot}$}
\def \kms {$\,$km s$^{-1}$}

\def \mJyb {$\,$mJy$\,$beam$^{-1}$}
\def \h{$^{\rm h}$}
\def \m{$^{\rm m}$}
\def\sec{\hbox{$^{\rm s}$}}
\def \ad {$\arcdeg$}
\def \am {$\arcmin$}
\def \as {\hbox{$\arcsec$}}
\def \HI {H{\small $\,$I}}

\begin{document}

\title{VLBA Observations of \HI~ in the Archetype Compact 
Symmetric Object B2352+495}

\author{E. D. Araya\altaffilmark{1,}\altaffilmark{2,}\altaffilmark{3},
C. Rodr\'{\i}guez\altaffilmark{2},
Y. Pihlstr\"om\altaffilmark{2,}\altaffilmark{4},
G. B. Taylor\altaffilmark{2,}\altaffilmark{4}, 
S. Tremblay\altaffilmark{2}
\and
R. C. Vermeulen\altaffilmark{5}}

\altaffiltext{1}{Jansky Fellow of the National Radio Astronomy 
Observatory, P.O. Box 0, Socorro, NM 87801.}

\altaffiltext{2}{Department of Physics and Astronomy,
MSC07 4220, University of New Mexico, Albuquerque, NM 87131.}

\altaffiltext{3}{Physics Department, Western Illinois
University, 1 University Circle, Macomb, IL 61455.}

\altaffiltext{4}{Adjunct Astronomer, National Radio Astronomy
Observatory.}

\altaffiltext{5}{ASTRON, Netherlands Foundation for Research 
in Astronomy, PO Box 2, 7990 AA Dwingeloo, The Netherlands.}

\begin{abstract}

B2352+495 is a prototypical example of a Compact Symmetric
Object (CSO). It has a double radio lobe symmetrically
located with respect to a central flat spectrum radio core
(the location of the AGN) and has a physical extent of
less than 200$\,$pc. In this work we report VLBA observation 
of 21$\,$cm \HI~absorption toward B2352+495 to investigate the properties
of this remarkable radio source, in particular, to explore
whether the radio emission can be confined by circumnuclear
material (frustration scenario) or whether the source
is likely to be young. We confirmed the two \HI~absorption 
features previously detected toward B2352+495
-- a broad line nearly centered at the systemic velocity of the
galaxy and a narrow redshifted component. The atomic
gas from the broad absorption component is likely associated
with circumnuclear material, consistent with the current 
paradigm of clumpy \HI~distribution in toroidal structures 
around supermassive black holes.
\end{abstract}


~
\pagebreak

\section{Introduction}

Compact Symmetric Objects (CSOs) are characterized by compact 
bright radio emission at scales $<1\,$kpc
with lobes on both sides of a 
central engine (Wilkinson et al.$\,$1994). CSOs tend
to be associated with bright elliptical galaxies, 
although some appear to be associated with quasars and
even spiral galaxies (Readhead et al.$\,$1996;
Augusto et al.$\,$2006; Perlman et al.$\,$1996). Many CSOs have an overall 
GigaHertz-Peaked Spectrum (GPS), and the cores are often 
undetected at frequencies of 5$\,$GHz or 
higher (Taylor et al.$\,$1996). CSOs are very luminous radio sources,
significantly brighter (by more than a factor of 10) than 
the luminosity threshold between FR$\,$I and
II radio galaxies (i.e., $\sim 2\times 10^{25}\,$W$\,$Hz$^{-1}$sr$^{-1}$
at 178$\,$MHz, Fanaroff \& Riley 1974; Readhead et al.$\,$1996). 

It has been proposed that the small source size could be
caused by ``frustration'' of the radio jets, which 
fail to escape from a dense galactic medium (e.g.,
van Breugel et al.$\,$1984; O'Dea et al.$\,$1991;
De Young 1993). However, a different scenario is currently preferred:
CSOs may be young progenitors ($\la$10$^3$ -- 10$^5\,$years) 
of powerful extended radio sources (e.g., Phillips \& Mutel 1982; 
Augusto et al.$\,$2006; Augusto 2007).
Based on XMM-Newton observations
of five CSOs, Vink et al.$\,$(2006) found that their absorption
column density ($N_H \sim 10^{22}\,$cm$^{-2}$) 
is on average greater than that
of broad-line radio galaxies, but similar to column densities
observed toward narrow-line and other types of radio galaxies (NLRGs, RG; see 
Sambruna et al.$\,$1999 for definitions). 
These results argue against the frustration scenario because much higher X-ray
absorption column densities than found by Vink et al. (2006) are
expected if the expansion of the radio jets were confined 
(see also Guainazzi et al. 2006).
In contrast, the dynamical age of CSOs measured from the 
expansion velocity of hotspots in the radio lobes 
supports the youth hypothesis
(Taylor et al.$\,$2000, Polatidis \& Conway 2003).
The youth interpretation is also supported by 
limits on molecular gas content (O'Dea et al.$\,$2005), 
\HI~observations (Pihlstr\"om et al.$\,$2003), and under-luminous 
O{$\,$\small III} line emission (Vink et al.$\,$2006).

The study of CSOs is therefore fundamental
for the understanding of the formation and evolution of 
radio galaxies. In addition, CSOs are good candidates
to investigate the fueling of supermassive black holes given that
the compact radio emission enables very high angular resolution
observations of circumnuclear gas via absorption studies.
In this work we report high sensitivity and high angular 
resolution observations of \HI~toward B2352+495; 
an archetype of the CSO class.

\subsection{\HI~Absorption in Compact Symmetric Objects}

According to Curran et al.$\,$(2008), \HI~absorption associated with
sources at redshifts greater than $\sim 0.1$ has been detected
toward 37 sources, 13 of them classified as CSOs, GPS, or
HFP (High Frequency Peaked) sources.
An overview of \HI~absorption in compact radio sources 
was published by Vermeulen et al.$\,$(2003a). They report an
\HI~detection rate of $33\%$ based on Westerbork
Synthesis Radio Telescope (WSRT) observations of 59
sources (i.e., 19 detections). Of the 59 sources, 11
were classified as CSO by Augusto et al.$\,$(2006) and Polatidis
\& Conway (2003). From this subsample of 11 CSOs, 6 (54$\%$) show 
\HI~absorption. Other studies show similar results, e.g., 
Peck \& Taylor (2001) report a CSO \HI~detection rate of $\sim 50\%$, 
and Gupta et al.$\,$(2006) report a detection rate of $\sim 45\%$
in GPS sources (see also Vermeulen 2002).

The optical depth range of the 6 CSOs with \HI~absorption 
from Vermeulen et al.$\,$(2003a) is 0.2 to 1.7$\%$, 
with linewidths between 13 and 297\kms~and
column densities between $2.8\times 10^{19}$ and 
$2.6\times 10^{20} \,$cm$^{-2}$.
To estimate \HI~column densities, Vermeulen et al.$\,$(2003a)
assumed that the atomic gas was uniformly covering the 
radio continuum (covering factor of 1) and a spin
temperature of 100$\,$K, that may underestimate the spin 
temperature by more than an order of magnitude if the 
gas were located at parsec-scales from the AGN. Thus,
the column densities reported by Vermeulen et al.$\,$(2003a)
are lower limits.

Including the results reported in this work, \HI~imaging 
at high angular resolution has been reported toward five CSOs 
(Table~1). We also list in Table~1 all non-CSO 
AGN and starbursts that have been imaged at high angular resolution
in \HI, i.e., in interferometric observations using baselines 
greater than 100$\,$km. In the rest of this section we
review the most salient characteristics of the CSOs that
have been studied in \HI~at high angular resolution:

\noindent {\it --- B1946+708:} High angular resolution 
\HI~observations were conducted by Peck et al.$\,$(1999; 
see also Peck \& Taylor 2001).
They found extended and multi-peaked absorption throughout 
the radio continuum source. The peak optical depths 
range between 3 and 7$\%$. The maximum column density 
($N_{HI} \sim 3 \times 10^{23}\,$cm$^{-2}$; Peck \& Taylor 2001)
was found toward the AGN core, which also shows the 
greatest velocity dispersion ($FWHM \sim 350$\kms).
The extent of the \HI~absorption in B1946+708 implies 
that the atomic gas is tracing a thick torus (thickness $> 80\,$pc).

\noindent {\it --- 4C$\,$31.04:} The host galaxy is a bright 
elliptical at z = 0.0592. It was classified as CSO 
based on VLBI continuum observations that revealed a
double radio lobe with a flat spectrum radio
core (Cotton et al.$\,$1995; Giovannini et al.$\,$
2001, Giroletti et al.$\,$2003; see also Augusto et al.$\,$1998).
Detection of \HI~in 4C$\,$31.04 was reported by Mirabel (1990) based on
Arecibo observations. Mirabel (1990) found two absorption features,
a broad (FWHM = 133\kms) component and a double peaked 
narrow component (FWHM = 6 and 16\kms) redshifted approximately 
400\kms~from the optical velocity.
Subsequent VLBI observations by Conway (1999) 
showed that the broad line covers one of the radio
lobes and just partially the second, indicating a 
sharp edge of the atomic gas torus.
The narrow component is detected toward both radio lobes. 
The maximum optical depth of the lines is $\sim 5$ to $7\%$.
Mirabel (1990) interpreted the narrow absorption as 
being caused by clouds similar to high velocity
\HI~clouds in our Galaxy. 

\noindent {\it --- 4C$\,$37.11:} In contrast to the 
two previous CSOs, \HI~in 4C$\,$37.11 has been detected only toward 
one of the radio lobes. The absorption is characterized by
two broad (FWHM$>$100\kms) \HI~features 
from two separated (in line-of-sight and velocity) clouds. 
The peak optical depth of the \HI~lines is $\sim 2\%$.
Rodr\'{\i}guez et al.$\,$(2009) 
concluded that the \HI~absorption likely originates in a
thick circumnuclear torus. 

\noindent {\it --- PKS$\,1413+135$:} As in 4C$\,$37.11, 
\HI~absorption is detected only
toward one radio lobe (Perlman et al.$\,$2002).
The \HI~optical depth is $\tau \approx 1$ and the line
is quite narrow, between 16 and 18\kms. Perlman
et al.$\,$(2002) concluded that the \HI~absorption is likely
from a giant molecular cloud in the outer disk of the host galaxy.
Although PKS$\,1413+135$ was classified by Perlman et al.$\,$(1996) 
and Gugliucci et al.$\,$(2005)
as a CSO, it has several characteristics that differ form
a typical CSO. In particular, it was originally classified 
as a BL Lac object based on its optical spectrum (Bregman et al. 1981),
the nucleus dominates by more than 50$\%$ of the total cm 
flux density, the radio emission is variable, and  
the host appears to be a spiral galaxy (see Perlman et al.$\,$1996).

Finally, two additional sources (B1934$-$638 and 4C$\,$12.50) are worth 
discussing in detail. B1934$-$638 is a compact-double GPS 
source that has been
classified as a CSO (Augusto et al. 2006), however, no flat-spectrum
radio core has been detected (Tzioumis et al. 2002). Even though
no \HI~imaging has been reported at high angular resolution,
the \HI~absorption is against the compact radio continuum as shown 
by detection of absorption with three baselines of the LBA (V\'eron-Cetty 
et al. 2000). The absorption is narrow (FWHM = 18\kms) and 
redshifted 260\kms~from the optical systemic velocity, thus, the \HI~is
probably not located in the nuclear region of the galaxy. 
We include B1934$-$638 in Table~1 although no VLBI 
\HI~imaging is available and the radio core has not been detected.

In the case of 4C$\,$12.50, high angular resolution \HI~absorption observations were
reported by Morganti et al.$\,$(2004). They found 
\HI~absorption against a
weak counterjet and concluded that the \HI~likely
traces a jet/cloud interaction and not a circumnuclear torus
or disk. This source was classified as a CSO by Lister et 
al.$\,$(2003) despite several atypical characteristics.  
For instance, while the 1.3$\,$GHz continuum image reveals 
a S-shaped distribution with a total extend of $\sim 300\,$pc
(Morganti et al.$\,$2004), the 15$\,$GHz VLBA observations
are consistent with a core-jet morphology plus a weak 
counterjet (i.e., inconsistent with the typical symmetrical 
morphology of CSOs). 4C$\,$12.50 has no clear hotspots, it is
classified as a CSS (Compact Steep Spectrum), the radio jet exhibits
superluminal motions, some isolated features have 
unusually high ($\sim 60\%$) linear polarization in comparison 
to other CSOs, and more importantly, there appears to be continuity between
the $\sim 300\,$pc compact radio jets and large scale ($>10\,$kpc)
radio lobes (Stanghellini et al.$\,$2005). 
Therefore, we do not consider 4C$\,$12.50 a bonafide member of the
CSO class (see Table~1).

\subsection{B2352+495: An Archetype Compact Symmetric Object}

B2352+495 has been studied in detail by a number of authors 
(e.g., Readhead et al.$\,$1996, Taylor et al.$\,$1996); in this
section we summarize the characteristics of this object:

\noindent {\it Radio Continuum:} The most salient property
of B2352+495 is its highly symmetric morphology at radio
wavelengths. It exhibits an S-shaped morphology with a total
extent of $\sim 120\,$pc (Readhead et al.$\,$1996). High
angular resolution radio continuum observations from 610$\,$MHz to 
15$\,$GHz reveal symmetric hotspots and lobes; remarkably
similar to the morphology of FR II objects (but at much smaller
scale). The dynamical age of the system is $\sim 1200\,$yr
based on proper motion studies of the hotspots in the radio lobes
(Taylor et al.$\,$2000).
Taylor et al.$\,$(1996) detected the radio core of the system, 
i.e., the putative location of the central engine. 
The core is approximately located symmetrically between the 
radio lobes, and
drives a collimated jet that is brighter toward the 
northern side of the core. The radio jet is detected up to 
the brightest radio emission in the system (B1 and B2; Taylor
et al 1996; see also $\S 3$) which is also located
between the two radio lobes. The hotspots in the radio 
lobes show no superluminal motion (Conway et al.$\,$1992).
Multi-frequency VLA observations give fractional polarization
limits of $<1\%$ for this source (Rudnick \& Jones 1983).
The radio spectral energy distribution of B2352+495
peaks at $\sim 1\,$GHz, thus by definition, it is a
GPS object (Conway et al.$\,$1992).
The total flux density of B2352+495 shows only modest
($\la 20\%$) variability on time scales of months; no
systematic long term variability was found over a 
period of $\sim 15\,$years (Waltman et al.$\,$1991;
Conway et al.$\,$1992). The radio luminosity is 
$\sim 10^{26}\,h^{-2}\,$W$\,$Hz$^{-1}$ ($\nu_o = $5$\,$GHz 
at the emitted frame), i.e., some 20 times
brighter than the luminosity boundary between FR Is
and FR IIs (Readhead et al.$\,$1996).

\noindent {\it Optical/Infrared Counterpart:} B2352+495 was not
detected by IRAS to sensitivity limits well below the corresponding
luminosity of Arp$\,$220 (at the redshifted frequency), thus 
B2352+495 is significantly less luminous than ultra-luminous
infrared galaxies (Readhead et al.$\,$1996). 
Snellen et al.$\,$(2003) conducted imaging and spectroscopic optical
observations of B2352+495.
They found that the radio source resides in an optical host in
the fundamental plane of elliptical galaxies, i.e., the host is a 
normal elliptical.$\,$Snellen et al.$\,$(2003) report a 
redshift
of 0.23790$\pm$0.00016, stellar velocity
dispersion of 201$\pm$17\kms, and an effective radius of 
1.6\arcsec$\pm$0.3\arcsec~(5.8$\pm$1.1$\,$kpc).\footnote{At 
a redshift of 0.23790,  
10$\,$mas correspond to 36.3$\,$pc assuming a cosmology of 
$H_0 = 73.0$\kms$\,$Mpc$^{-1}$, $\Omega_{matter} = 0.27$, 
$\Omega_{vacuum} = 0.73$.} Based on the stellar velocity dispersion,
the mass of the supermassive black hole at the center of B2352+495
is of the order of $10^8$\Mo. The absolute magnitude of the galaxy is
$M_V = -19.8$ with an apparent magnitude of $m_V = 20.1$ 
(Readhead et al.$\,$1996). The optical luminosity of B2352+495 is
$\sim 0.35\,L^*$, which is smaller
than the typical optical luminosity of Classical Doubles (FR IIs; 
$L \sim L^*$; Owen \& White 1991) and cD galaxies 
($L \sim 5 - 10\,L^*$; Schechter 1976). The absence of Balmer lines
and strong blue continuum in the spectrum demonstrates no
starburst activity. In contrast, a number of emission lines
are detected typical of active elliptical galaxies (Readhead et al.$\,$1996). 
The AGN contribution to the total optical continuum 
is $\la 10\%$ (Snellen et al.$\,$2003).

\noindent {\it X-Ray Properties:} 
High sensitivity XMM-Newton observations of B2352+495 were conducted by Vink 
et al.$\,$(2006). They detected an X-ray source with a luminosity 
of $4.6 \times 10^{42}\,$erg$\,$s$^{-1}$ ($2 - 10\,$keV band, ignoring 
absorption). Based on the X-ray detection, they estimated an intrinsic
absorption column density of (0.66$\pm$0.27)$\times 10^{22}\,$cm$^{-2}$, 
which is similar to the column densities found toward extended radio
galaxies.

\noindent {\it Prior \HI~Observations: }
Absorption of the \HI~transition in B2352+495 was first detected
by Vermeulen et al.$\,$(2003a) based on
WSRT observations. 
The \HI~absorption in B2352+495 consists of a broad line (FWHM = 82\kms)
and a redshifted narrow component (FWHM = 13\kms).
The peak flux density of the two absorption components is 
similar ($\sim 40\,$mJy). Preliminary results of the data reported
in this paper were discussed by Vermeulen (2002).

\section{Observations}

Very long baseline interferometer (VLBI) 
observations were conducted on December 01 and 02, 1997, with
the NRAO Very Long Baseline Array (VLBA\footnote{The VLBA is operated
by the National Radio Astronomy Observatory (NRAO), a facility of the 
National Science Foundation operated under cooperative agreement by 
Associated Universities, Inc.}) and the WSRT. 
The data were correlated with the VLBA correlator at the
P.V.D. Science Operations Center in Socorro, New Mexico.
The target source (B2352+495) was observed during approximately
8.5$\,$hours each day ($\sim 17\,$hours total) in the 21$\,$cm 
\HI~transition ($\nu_0 = 1420.406\,$MHz) which is 
redshifted to a sky frequency of $\nu_{sky} = 1147.40\,$MHz
(z = 0.23790$\pm$0.00016, Snellen et al.$\,$2003; 
V$_{Hel}$ = 71321$\pm$48\kms, optical definition).
We observed with 
a bandwidth of 8$\,$MHz, 
1024 channels, dual circular
polarization, and an initial channel width of $\Delta \nu = 7.8\,$kHz
smoothed to a spectral resolution of 15.6$\,$kHz (4.1\kms). 
Only eight of the ten VLBA antennas were used because of RFI filters
that block access at Fort Davis and Kitt Peak. 
We observed 3C84 for bandpass calibration.

All data reduction was done in the NRAO package AIPS using standard
VLBA spectral-line calibration procedures. The WSRT data were
significantly affected by RFI and we decided not to use them
after careful inspection of the results obtained
including and excluding WSRT.
We used line-free channels of the B2352+495 observations for 
fringe fitting. A continuum 
data set was obtained by averaging the line-free channels
of the visibility file; the continuum data were used for self-calibration. 
The absolute astrometry of the data was therefore lost, 
and we simply assign the radio continuum peak 
to the phase tracking center (RA = 22\h55\m09.4581\sec, 
Decl. = +49\ad50\am08.340\as, J2000).

Given the non-standard frequency of the observations, a reliable flux
density calibration using the T$_{sys}$ method was not possible,
thus we scaled the flux density to a total radio continuum
emission of 2.55$\,$Jy at 1.14$\,$GHz. This value was obtained by 
interpolating the flux density measurements reported by Readhead et al.$\,$
(1996). The use of the Readhead et al.$\,$(1996) values to interpolate
the flux density is valid because the radio emission is dominated
by extended radio lobes (see below) and not by a compact core
(and thus it is less likely to exhibit strong variability). 
Despite the uncertainty on the flux density calibration,
the peak flux density values of the narrow and
broad \HI~lines detected in this work are very similar
to those obtained with WSRT (Vermeulen et al.$\,$2003a). 
In addition, our discussion of \HI~absorption is based on 
opacity considerations, which are not affected by systematic flux
density calibration errors. 

We transferred the self-calibration
tables from the continuum data set to the line file, and subtracted 
the radio continuum using the task UVLIN in AIPS. We used a Brigg's 
robust 0 weighting for the final imaging that resulted in a 
$\theta_{syn} = 8.3 \times 6.7\,$mas ($\sim 27\,$pc 
linear scale),
PA = 0.5\arcdeg, rms = 0.71\mJyb~(in continuum) 
and 7.1\mJyb~(per frequency channel). All velocities reported in the
paper are in the rest frame of the CSO (assuming $z=0.23790$) unless 
indicated otherwise.

\section{Results}

We detected both radio continuum and compact \HI~absorption in B2352+495.
In Figure~1 we show the \HI~spectrum (top panel); the bottom panel
shows the 1.14$\,$GHz continuum and the zero velocity moment
(integrated intensity) map of the \HI~detection.
The radio continuum has a symmetric S-shaped distribution that
is consistent with previous radio continuum images of the source
(e.g., Wilkinson et al.$\,$1994, Readhead et al.$\,$1996, Owsianik et al.$\,$1999).
\HI~absorption is only detected toward the brightest continuum emission
close to the center of the CSO. In Figure~2 we show the optical depth
channel maps; only signal above 4$\sigma$ in the continuum
and \HI~data were used to calculate the opacity.

The \HI~absorption spectrum has two features,
a broad component 
($\Delta V_{>1\sigma} = 102 \pm 8$\kms, FWHM $= 85 \pm 4$\kms, 
$V_{peak} = -9.9 \pm 1.6$\kms) 
that is almost centered at the systemic velocity of the 
galaxy (\HI~centroid Heliocentric velocity = $71307 \pm 10$\kms; 
systemic velocity V$_{Hel}$ = 71321$\pm$48\kms, optical definition), 
and a narrow component (FWHM $= 13\pm 2$\kms)
redshifted from systemic by $V_{peak} = 129.9 \pm 0.8$\kms.
Our measurements are therefore consistent with those of 
Vermeulen et al.$\,$(2003a) within our velocity resolution.\footnote{We note
that there is a typo in the sky frequency of the narrow HI line in
B2352+495 reported by Vermeulen et al.$\,$(2003a), it should read 1147.0 
instead of 1147.2$\,$MHz.}

\section{Discussion}

As in the case of the \HI~gas associated with the nucleus of
B1946+708 (Peck et al.$\,$1999), we estimate the \HI~column density
assuming a covering factor of 1 and
a spin temperature of 8000$\,$K, i.e., assuming that the \HI~gas 
is directly associated with the AGN.
Using the standard column density/opacity 
relation (e.g., Rohlfs \& Wilson 2000), we estimate
a total column density of $(5.2\pm0.2) \times 10^{22}$ and
$(7.3 \pm 1.0) \times 10^{21}\,$cm$^{-2}$ for the broad and narrow
\HI~lines, respectively (only statistical errors 
from the fit were used to estimate the column density uncertainty). 
The values are greater than the B2352+495 \HI~column densities 
reported by Vermeulen et al.$\,$(2003a), however,
as mentioned above, their values are lower limits because
of the assumption of uniform coverage of the radio continuum
(which is not the case, see Figure~1) and the 
assumption of a lower spin temperature ($T_{sp} = 100\,$K).
If the \HI~clouds have physical dimensions comparable
to the synthesized beam (note that there appears to be structure
in the opacity channels, Figure~2), then the \HI~density would be 
between $\sim 10^2$ and 10$^3\,$cm$^{-3}$, corresponding to total 
\HI~masses of
$\sim 2\times10^5$ and $3 \times 10^4\,$\Mo~for the broad and narrow
features, respectively. 

If the \HI~absorption profile of the broad component is the
superposition of many narrow \HI~features, and assuming 
a typical optical depth of $\tau = 0.04$ for the narrow
features, then we could not have detected \HI~absorption 
(at a $\ge$3$\sigma$ level)
anywhere except toward the brightest 
continuum region at the center of the CSO (i.e., the region where
\HI~absorption was indeed detected). This implies that, given our
sensitivity limit, our data do not set significant constraints
on the scale height of the \HI~distribution.

As discussed below, we favor the interpretation that
the broad \HI~component is associated with rotation of a
disk/torus around the supermassive black hole in B2352+495.
The detection of a narrow \HI~feature redshifted with respect 
to the systemic velocity indicates infall.
However, we cannot establish whether 
the \HI~feature is associated with circumnuclear material
or with a foreground \HI~cloud in the host galaxy, e.g,  
whether it is located in the narrow-line region (NLR; e.g., B2050+364, 
Vermeulen et al.$\,$2006) or 
associated with a high velocity cloud.
We note that other studies
(e.g., van Gorkom et al.$\,$1989, Peck \& Taylor 1998, 
Conway 1999) have also found evidence for infalling material 
in galaxies with compact radio cores (see however Vermeulen et al.$\,$2003a).

We now discuss the nature of the radio continuum having the \HI~absorption.
In principle, the brightest 1.14$\,$GHz continuum source
(located close to the center of the two radio lobes, Figure~1) 
is suggestive of the active core, i.e., the location of the 
supermassive black hole. However, comparing our continuum
data with higher angular resolution observations reveal
a different scenario. 
In Figure~3 (upper panel) we show our 1.14$\,$GHz continuum map
superimposed on the 15.4$\,$GHz observations of Taylor 
et al.$\,$(1996). Both continuum data sets were self-calibrated, 
and thus no absolute astrometry is available. 
To overlay the two images, we
used an average offset computed from the peak in the northern and the
southern radio lobe emission; we assume that position shifts due to
opacity effects are small there, since their measured separation is the
same within the errors at 1.14$\,$GHz and 15.4$\,$GHz.
The brightest  
radio emission at the center of the radio lobes was {\it not} used to 
estimate the offset.

Taylor et al.$\,$(1996) found a compact radio source characterized by
an inverted spectral index ($\alpha = 0.14\pm0.3$; $S_\nu \propto \nu^\alpha$), 
which convincingly indicates the 
location of the active core (see Figure~3). 
Assuming a flat spectral index between 1 and 8$\,$GHz, 
the flux density of the core at 1$\,$GHz is $\sim 13\,$mJy.
Our \HI~optical depth limit toward the radio core is $<$0.1 (3$\sigma$). 
A prominent radio jet is detected to the north of the core, and a 
weaker counterjet to the south (Taylor et al.$\,$1996; 2000). 
The northern edge of the jet is coincident with
the brightest 1.14$\,$GHz continuum source 
reported in this work, implying that
the \HI~absorption is not against continuum from the core, but
radio emission from the jet. Among other possibilities, 
the detection of the radio jet and weaker counterjet could 
be due to relativistic
beaming (see Taylor et al.$\,$[1996] for this and other interpretations).
The \HI~absorption in B2352+495 is thus similar to the absorption 
observed in B0402+379 (Rodr\'{\i}guez et al.$\,$2009) and
NGC$\,$4151 (Mundell et al.$\,$1995, 2003)
where \HI~is detected only toward one side of the radio jet. 

It is interesting to note that the radio continuum source toward which
the \HI~is detected (Figure~3) is the location where the radio 
jet begins to change its position angle. This would suggest that, as 
for example in the case of 
3C$\,$236 (Conway 1999), the \HI~absorption marks the location
of a jet/cloud interaction responsible for disturbing (and possibly bending) 
the jet. However, the B2352$+$495 morphology is quite
symmetric in both radio lobes which suggests that the deviation
of the radio jets is not due to local jet/cloud interactions but 
governed by the large scale magnetohydrodynamical properties of the
galactic nucleus. In the next section, we propose a different 
interpretation for the nature of the \HI~absorption in B2352+495.

\subsection{Possible \HI~Velocity Gradient and Disk/Torus Interpretation}

It is difficult to reliably 
investigate the velocity field of the broad \HI~line
given that the absorption is only detected toward the 
central continuum source and, as shown in Figure~2, 
the velocity field is complex. Nevertheless, a
velocity gradient may be present: the two channels with stronger 
absorption at velocities above 10\kms~from systemic
appear to have stronger absorption toward the west, whereas
most of the blueshifted absorption is stronger toward the east 
(velocity channels $< -10$\kms~from systemic, see
Figure~2). Figure~4 shows the position-velocity 
diagram of the broad \HI~absorption line in the 
east-west direction. As also seen in the individual channel
maps, there is a hint of a velocity gradient between
blueshifted and redshifted material. 
To illustrate the possible gradient, we show in 
Figure~3 (top panel) the opacity
channels that correspond to the rest frame velocities of 
17.6 and $-31.4$\kms~(Fig.~2). 

The existence of a velocity gradient is tantalizing given
that it is approximately centered at the
systemic velocity of the galaxy, and the position angle of
the gradient is almost perpendicular to the 
north-south jet, i.e., the \HI~gas could be 
tracing some kind of rotating structure
around the supermassive black hole. 
We explore this hypothesis in the rest of this 
section, however, we stress that more sensitive 
global-VLBI observations are required to confirm 
the velocity gradient.

To investigate whether rotation around the supermassive
back hole is physically plausible, we followed the formulation 
presented by Rodr\'{\i}guez et al.$\,$(2009; see their
appendix). To model the \HI~rotating structure, we used the position
and velocity of the peak optical depth of the 17.6 
and $-$31.4\kms~channels (Figure~2). 
The position of the radio core (Figure~3 upper panel; see also Taylor et al.$\,$1996)
was assumed to be coincident with the center of mass
of the system. For a black hole mass of 
10$^8$\Mo~(Snellen et al.$\,$2003) and assuming circular 
trajectories coplanar with the supermassive black hole, 
we obtain that the orbit of the \HI~clouds
from the 17.6 and $-$31.4\kms~channels would have a
radius of $\sim 20\,$pc with an orbital inclination of 
$\sim 65$\arcdeg~with respect to the plane of the sky.
Such an orbit is reasonable given the expected distribution of
atomic clouds around AGNs (e.g., Peck 2005).
Similar results are obtained in the case that the orbital plane
of the \HI~clouds is not coplanar with the supermassive
black hole but located above by distances of the order of the projected
separation between the AGN core and the \HI~absorption.
This estimate is clearly a zero-order approximation
because non-radial inhomogeneities in the gravitational
field due to stars and gas are neglected. In addition, the radius of influence
of the supermassive black hole ($r_{BH} = G M_{BH} / \sigma_*^2$, where
$G$ is the gravitational constant, $M_{BH}$ is the mass of the
black hole, and $\sigma_*$ is the stellar velocity dispersion; 
e.g., Neumayer et al. 2007) is $\sim 10\,$pc, thus, the 
stellar mass enclosed by the atomic gas is likely comparable to the
black hole mass.

In Figure~3 (bottom panel) we show two
possibilities for the nature of the \HI~distribution, 
i.e., a thick torus (characterized by an \HI~scale height
similar or greater than the length of the radio source)
or a thin (most likely warped, e.g., Pringle 1996) disk.
The obvious method to discriminate between these
possibilities is to check whether there is \HI~toward the southern
radio lobe. However, as mentioned above, our optical depth 
sensitivity limit does not set
significant constraints on the \HI~distribution, i.e., 
clouds like those detected toward the brightest radio
continuum emission could be present throughout the
region. 

An indirect approach to discriminate between the models is
to compare our observations with lower angular 
resolution data.
As mentioned in $\S 2$, the peak flux density
measured with WSRT (Vermeulen et al.$\,$2003a) is 
approximately the same peak intensity we measured with the 
VLBA (Figure~1) and the linewidths are also consistent.
Thus, in contrast to other systems where VLBI observations resolve/filter
out significant \HI~signal (e.g., IC$\,$5063, Oosterloo et al.$\,$2000), 
the \HI~in B2352+495 is compact and mostly distributed 
along the central continuum
source, i.e., consistent with the thin disk interpretation.

The thin disk interpretation is also circumstantially supported
by X-ray observations. Vink et al.$\,$(2006) reported
a X-ray absorption column density of 
$(6.6\pm2.7) \times 10^{21}\,$cm$^{-2}$,
which is similar to the column density of
the narrow \HI~component, but several times smaller than 
the value of the broad line. If the \HI~absorption originates from 
an inclined (and warped) thin disk that does not substantially
obscure the AGN (Figure~3, bottom panel; see
also Fig. 5 of Orienti et al.$\,$2006) then as observed, 
the column density toward the AGN derived from X-ray 
observations is expected to be smaller than that of the neutral disk.

We note that the thin disk  
interpretation does not necessarily imply a misalignment
between the jet and the disk, i.e., 
the position angle of the jet changes
as a function of distance from the core (Figure~3) and
the disk could be warped, thus, the jet and disk could be 
perpendicular at (sub)parsec scales. In addition, given that the detection
rate of \HI~absorption in CSOs is $\sim 50\%$ (see $\S 1.1$),
it is certainly possible that in some cases the atomic gas is
distributed in a circumnuclear structure that is seen in
projection only toward one radio lobe/jet.

Assuming the orbital parameters from the thin disk model,
we now explore the stability of the \HI~disk with respect to 
gravitational fragmentation. We used the Toomre stability 
formulation as presented by Gallimore et al.$\,$(1999), i.e., 

\begin{equation}
\frac{\Sigma_{HI}}{\Sigma_c} \simeq \frac{0.028 N_{HI} v_{rot}}{r_{pc}},
\end{equation}

\noindent where $\Sigma_{HI}$ is the atomic hydrogen surface density,
$\Sigma_c$ is the critical surface density above which the disk
is unstable to fragmentation, $N_{HI}$ is the \HI~column density in
units of 10$^{21}\,$cm$^{-2}$, $v_{rot}$ is the rotation velocity
in \kms, and $r_{pc}$ is the orbital radius in parsecs. In the case of 
B2352+495 we obtain $\Sigma_{HI}/\Sigma_{c} \sim 10$, thus, the disk
is expected to be unstable to fragmentation. We note that
reducing the assumed spin temperature by an order of magnitude would not
significantly modify this conclusion, i.e., 
$\Sigma_{HI}/\Sigma_{c}$ would still
be $\sim 1$. This simple stability test indicates that the \HI~cannot
be distributed in a homogeneous disk. Indeed, the opacity variations
from channel to channel (Figure~2) suggests that the \HI~gas
is not homogeneously distributed. We therefore concur with the
standard schematic of clumpy atomic gas around AGNs
(e.g., Peck et al.$\,$1999).

\subsection {Comparison with other High Angular Resolution Studies}

In this section we discuss the properties of \HI~absorption in 
B2352+495 with respect to other \HI~extragalactic absorbers 
that have been studied at very high angular resolution (i.e., interferometric
observations with baselines $> 100\,$km; Table~1).
As is clear from Table~1 and the discussion in $\S 1.1$, 
high angular resolution observations of \HI~in CSOs reveal 
that the absorption does not trace a specific kind of cloud, 
instead, the absorbing clouds appear to be
in quite different locations throughout the host galaxy, from infalling
high velocity clouds and giant molecular clouds at kiloparsec scales, 
to torus and disks closer ($\la 100\,$pc) to the nucleus. 

Only two of the five CSOs imaged at high angular resolution 
exhibit \HI~absorption toward both radio lobes (Table~1; B1934$-$638 
has not be imaged at high angular resolution, see $\S$1.1).
In general (four out of six, Table~1), CSOs studied at
high angular resolution are characterized by a broad ($\Delta V \ga$ 100\kms)
\HI~absorption component, and tend to be multi-peaked, with narrow
($\Delta V < $ 100\kms) lines detected in five of six cases. 

High angular resolution observations of \HI~toward non-CSO 
radio galaxies (Table~1) show a similar trend as
the CSOs counterparts, i.e., the \HI~absorption appears to trace
a variety of environments, from disks directly associated with 
the AGN to clouds at kiloparsec-scale distances from the active nucleus.
An apparent difference is the occurrence of \HI~clouds
interacting with radio jets. In the case of several non-CSO
with double jet/lobe radio emission, the \HI~appears to 
trace regions of interaction between atomic gas and radio jets, 
whereas such an interpretation has not been preferred in 
any of the CSOs imaged in \HI~at high angular resolution. 
This trend does not appear to be due to 
bias of the researchers but, 
in most cases, seems due to real differences in
the samples. For instance, the \HI~absorption 
associated with tori/disks in CSOs is found very close ($\la 100\,$pc) to the 
radio core, close to the systemic velocity, and/or has evidence
of rotation. Conversely, the \HI~in non-CSO double jet/lobe 
sources with possible
jet/cloud iteration is found at larger distances from the core
($\ga500\,$pc), shows no evidence of rotation, and/or has significantly 
different velocity from systemic.
Although a more robust statistical sample is
needed to achieve strong conclusions, the apparent 
absence of \HI~interaction with the radio jets supports the 
current view that CSOs are young radio galaxies instead of being
confined by dense circumnuclear material as proposed in the
frustration scenario.

Arguments for the youth scenario are mostly based on
the overall radio continuum properties of the sources,
power budget estimates of the jets, expansion velocities, and
properties of the nuclear molecular/atomic gas.
CSOs lack strong extended radio emission
related to energy deposited by the jets over long periods, 
and VLBI studies reveal proper motions that imply short
dynamic ages (of the order of $10^3\,$yr, e.g., Gugliucci et al. 2005). 
In addition, if the advance of the radio lobes would be truly frustrated, 
then for them to be in ram-pressure equilibrium the density of the 
surrounding medium should be above $10^2\,$cm$^{-3}$. However,
if the high density material is confined to a disk/torus
then the frustration scenario is not viable. 
B2352+495 has evidence of being a young source based on all these
arguments: it has no detected extended radio emission,
the expansion velocity implies a dynamic age of $\sim 1200\,$yr
(Taylor et al.$\,$2000), and, as discussed above,
the high density atomic gas appears to be distributed along a 
disk or torus which does not interact with the radio jet.
Hence, the evidence favors the youth
interpretation in this case.

\section{Summary}

We report high angular resolution ($\theta_{syn} \sim 8\,$mas)
observations of \HI~in the CSO B2352+495. We detect 
two absorption features: a broad line almost centered at the
systemic velocity of the host galaxy, and a redshifted 
narrow line indicating infall. Both lines are seen in 
absorption against the radio continuum of the central 
radio source of the CSO, which is the brightest continuum
source in the region at the observed frequency. The radio
continuum originates from the radio jet and not from the core.

In the case of the narrow \HI~component,
our data cannot establish the location of the \HI~cloud
with respect to the AGN, i.e., the radial distance of the cloud
to the galaxy center is unknown. On the other hand, the broad
component is likely tracing circumnuclear material
close ($<100\,$pc) to the AGN. 
Even though we cannot constrain the scale height of the \HI~gas,  
the broad \HI~absorption could originate from a 
clumpy circumnuclear disk.

All available data suggest that B2352+495 is a young radio
source and not an older source whose radio lobes have been
confined by high density circumnuclear material.$\,$Hence, this work
supports the study of CSOs as a key to understand the 
origin of extended radio galaxies.

\acknowledgments

We acknowledge an anonymous referee for critical 
comments that significantly improved the manuscript.
E.D.A. was partially supported by a postdoctoral fellowship at 
the University of New Mexico.
This research has made use of NASA's Astrophysics Data System,
and the NASA/IPAC Extragalactic Database (NED) which is operated 
by the Jet Propulsion Laboratory, California Institute of Technology, 
under contract with NASA.

{\it Facilities:} \facility{VLBA}

\clearpage

\begin{deluxetable}{lllll}
\tabletypesize{\scriptsize}
\tablecaption{\HI~Observations of AGN and Starbursts at High Angular Resolution$^*$}
\tablewidth{0pt}
\tablehead{
\colhead{Galaxy} &
\colhead{HI Detection$^a$}  & \colhead{Line Profile$^{b}$} & \colhead{HI Location$^c$}
& \colhead{Ref.} }
\startdata
\multicolumn{5}{c}{\it 1. Compact Symmetric Objects} \\
~\\
4C 31.04, B0116+319, J0119+3210           & Both lobes      & N, B, M & AGN torus +               & 1    \\
                                          &                 &         & infalling HVC             &      \\
4C 37.11, B0402+379, J0405+3803           & One lobe        & B, M    & AGN torus                 & 2    \\
OQ +122, PKS 1413+135, J1415+1320         & One lobe        & N       & GMC in kpc disk           & 3    \\
B1934$-$638, PKS 1934$-$63, J1939$-$6342  & \nodata$^{\ddagger}$  & N & Infalling HVC?            & 4    \\
B1946+708, J1945+7055                     & Both lobes      & N, B, M & AGN torus                 & 5,6  \\
B2352+495, J2355+4950                     & One lobe        & N, B, M & AGN disk +                & This \\
                                          &                 &         & infalling cloud           & work \\
~\\
\multicolumn{5}{c} {\it 2. Non-CSO Radio Sources: Double Jet/Lobe Systems as Imaged by VLBI}\\
~\\
3C$\,$49, B0138+136, J0141+1353                 & One lobe        & N, M    & Cloud/jet interaction     & 7    \\
NGC$\,$1052, B0238$-$084, J0241$-$0815          & Both radio jets & N, M    & Nuclear and/or galactic?  & 8    \\
Mrk 6, IRAS 06457+7429, J0652+7425              & One side of jet & N       & kpc disk                  & 9    \\
Hydra A, 3C$\,$218, B0915$-$118, J0918$-$1205   & Both jets+core  & N, M    & AGN disk                  & 10   \\
3C$\,$236, B1003+351, J1006+3454                & One lobe        & B       & Cloud/jet interaction     & 1    \\
NGC$\,$3894, B1146+596, J1148+5924              & Lobes + core    & N, M    & AGN torus                 & 11   \\
3C$\,$268.3, B1203+6430, J1206+6413             & One lobe        & N       & Cloud/jet interaction     & 7    \\
NGC$\,$4151, B1208+396, J1210+394               & One radio jet   & N, M    & AGN torus                 & 12   \\
NGC$\,$4261, B1216+061, J1219+0549              & Counterjet+core?& N, M?   & AGN disk                  & 13   \\ 
Mrk 231, IRAS 12540+5708, J1256+5652            & Diffuse cont.   & B       & $\sim 100\,$pc disk       & 14   \\
4C 12.50, IRAS 13451+1232, J1347+1217$^\dag$    & Counterjet      & B, M    & Cloud/jet interaction     & 15   \\
3C$\,$293, B1350+3141, J1352+3126               & Both jets+core  & B, N, M & Outer (8$\,$kpc) and      & 16   \\
                                                &                 &         & $<400\,$pc disk           &      \\ 
NGC$\,$5793, IRAS 14566$-$1629, J1459$-$1641    & Both lobes      & N, M    & $\sim$1$\,$kpc torus?     & 17   \\
NGC$\,$5929, B1524+4151, J1526+4140             & One lobe        & N       & AGN ring, or bar?         & 18   \\ 
PKS$\,$1814$-$63, J1820$-$6343                  & Both lobes      & M, N    & AGN disk +                & 19   \\
                                                &                 &         & Cloud/jet interaction?    &      \\
IC 5063, B2048$-$572, J2052$-$5704              & One lobe        & B, M    & Cloud/jet interaction     & 20   \\ 
TXS 2226$-$184, IRAS 22265$-$1826, J2229$-$1810 & Radio jet       & B       & Nuclear region,           & 21   \\
                                                &                 &         & cloud/jet interaction?    &      \\
NGC$\,$7469, B2300+0836, J2303+0852             & Radio core      & N       & Nuclear or galactic?      & 22   \\
PKS 2322$-$123, J2325-1207                      & One jet + core  & B, M    & AGN disk + infall?        & 23   \\ 
NGC$\,$7674, IRAS 23254+0830, J2327+0846        & One lobe + core?& B, N, M & AGN disk/torus            & 24   \\
~\\
\multicolumn{5}{c} {\it 3. Non-CSO Radio Sources: Single Jet + Core Systems as Imaged by VLBI}\\
~\\
NGC$\,$315, B0055+300, J0057+3021               & One jet + core  & N       & Infalling cloud           & 25   \\
NGC$\,$3079, IRAS 09585+5555, J1001+5540        & One jet + core  & N,M     & AGN torus                 & 26   \\
Cen A, NGC$\,$5128, B1322$-$428, J1325$-$4301   & One jet         & N, M    & kpc disk or rings         & 25   \\
PKS 1549$-$79, B1549$-$790, J1557$-$7913        & One jet + core  & N       & Nuclear region?           & 27   \\
Cygnus A, IRAS 19577+4035, J1959+4044           & Counterjet + core?& B     & Torus or 2$\,$kpc ring    & 1    \\
DA 529, B2050+364, J2052+3635                   & One jet + core  & N, M    & NLR/BLG ($<1\,$kpc)       & 28   \\
~\\
\multicolumn{5}{c} {\it 4. Non-CSO Radio Sources: Starbursts}\\
~\\
NGC$\,$2146, 4C+78.06, B06106+7822, J0618+7821  & Starburst cont. & B, N, M & Disk or ring              & 29   \\
IC$\,$694, Arp$\,$299, Mrk 171, B1125+588, J1128+5834    & Starburst cont. &  \nodata       & 250$\,$pc disk   & 30   \\
Mrk 273, B1342+5608, J1344+5553                 & Starburst cont. & B, M    & Starburst cores           & 31   \\ 
Arp$\,$220, B1532+2339, J1534+2330              & Starburst cont. & N, B, M & Disks in starburst cores  & 32   \\ 
PGC 060189, IRAS 17208$-$0014, J1723$-$0016     & Starburst cont. & B, N, M & Nuclear $\sim 1\,$kpc     & 33   \\   
\enddata
\tablecomments{$^{*}$ Interferometric observations with baselines greater than 100$\,$km.
The sources are divided in four groups, CSOs, non-CSO radio sources with detection of
two radio jets or lobes in opposite sides of a central core, non-CSO radio sources dominated 
by core-jet emission (although a weak counterjet may be detected, e.g., Cygnus A, Conway 1999), 
and starburst galaxies. 
$^{(a)}$ Indicates whether HI absorption was detected
towards all radio continuum emission (both lobes and core, or
starburst), or only towards one radio lobe.
$^{(b)}$ Detection of broad lines ($\Delta V > 100$\kms) is indicated with
`B', narrow lines ($\Delta V < 100$\kms) with `N', multiple lines in the 
spectrum (overlapped or otherwise) with `M'.
$^{(c)}$ Most likely (or at least preferred) location of the 
HI material with respect to the 
AGN as discussed in the reference paper, e.g., associated with a disk
of torus within $\sim$100$\,$pc from the AGN, with giant molecular 
clouds (GMC) or high velocity clouds (HVC) at galactic (kiloparsec) scales.
$^{(\ddagger)}$ The HI distribution in this source has not been imaged 
with VLBI, 
however, the absorption is against the compact radio continuum given 
detection of HI with three LBA baselines (V\'eron-Cetty et al. 2000; see also $\S1.1$).
$^{(\dag)}$ The classification of 4C$\,$+12.50 is controversial, see $\S1.1$.     
--- References: 
1. Conway (1999),
2. Rodr\'{\i}guez et al.$\,$(2009),
3. Perlman et al.$\,$(2002),
4. V\'eron-Cetty et al. (2000),
5. Peck et al.$\,$(1999), 
6. Peck \& Taylor (2001),
7. Labiano et al.$\,$(2006),
8. Vermeulen et al.$\,$(2003b), 
9. Gallimore et al.$\,$(1998),
10. Taylor (1996),
11. Peck \& Taylor (1998),
12. Mundell et al.$\,$(2003),
13. van Langevelde et al.$\,$(2000),
14. Carilli et al.$\,$(1998),
15. Morganti et al.$\,$(2004),
16. Beswick et al.$\,$(2004),
17. Pihlstr\"om et al.$\,$(1999),
18. Cole et al.$\,$(1998),
19. Morganti et al.$\,$(2000),
20. Oosterloo et al.$\,$(2000),
21. Taylor et al.$\,$(2004),
22. Beswick et al.$\,$(2002),
23. Taylor et al.$\,$(1999),
24. Momjian et al.$\,$(2003a),
25. Peck (1999; see also Morganti 2004), 
26. Sawada-Satoh, et al.$\,$(2000),
27. Holt et al.$\,$(2006),
28. Vermeulen et al.$\,$(2006),
29. Tarchi et al.$\,$(2004),
30. Polatidis \& Aalto (2001),
31. Carilli \& Taylor (2000), 
32. Mundell et al.$\,$(2001),
33. Momjian et al.$\,$(2003b).
}
\end{deluxetable}

\clearpage

\begin{figure}
\includegraphics{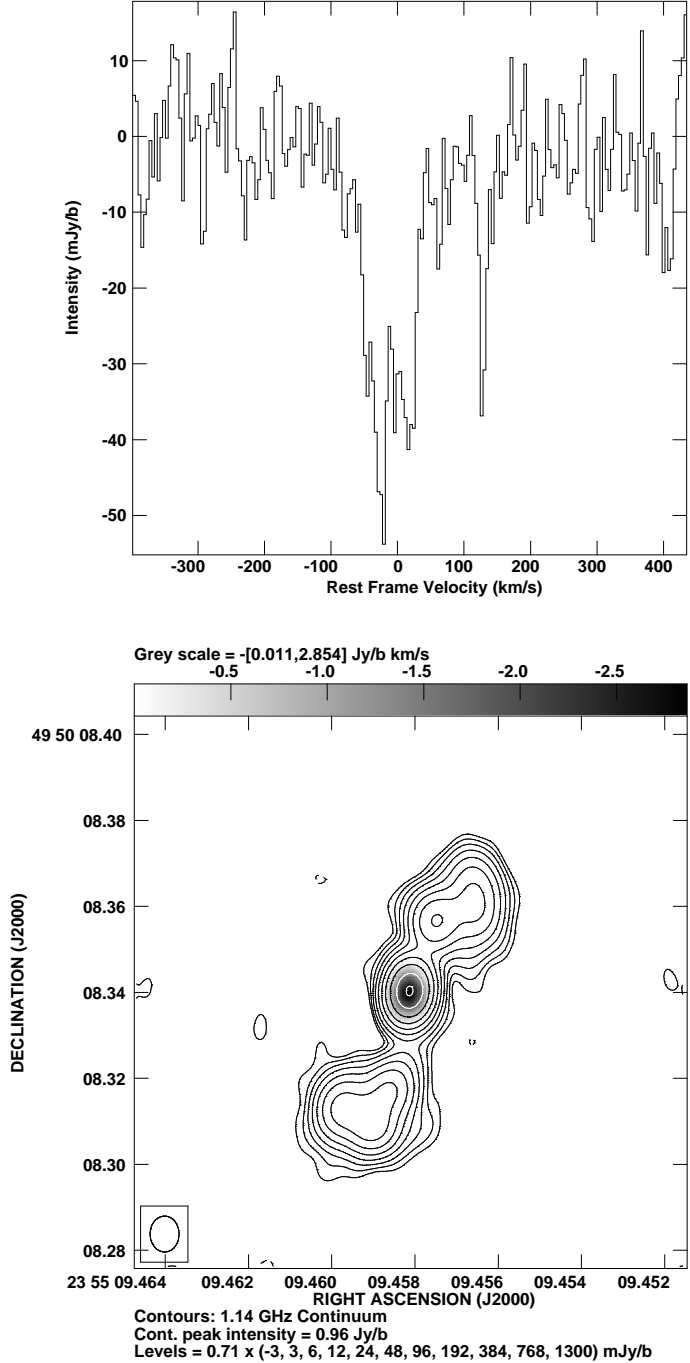} 
\vspace*{18.1cm}\caption{{\it Top:} \HI~absorption detected toward the compact
symmetric object B2352+495 after continuum subtraction. 
The systemic velocity of the system
(V$_{Hel}$ = 71321$\pm$48\kms, $V = 0.0$\kms~at the rest frame of the 
galaxy) is almost at the centroid of the broad absorption feature. 
{\it Bottom:} Zero velocity moment
(integrated intensity) of the \HI~absorption (grey scale) and 1.14$\,$GHz
continuum (contours). The data were self-calibrated and thus have no
absolute astrometry; the radio continuum peak was assigned
to the phase tracking center (RA = 22\h55\m09.4581\sec, 
Decl. = +49\ad50\am08.340\as, J2000). There is no 
registration difference between the continuum and \HI~absorption.}
\label{f1}
\end{figure}

\clearpage

\begin{figure}
\includegraphics{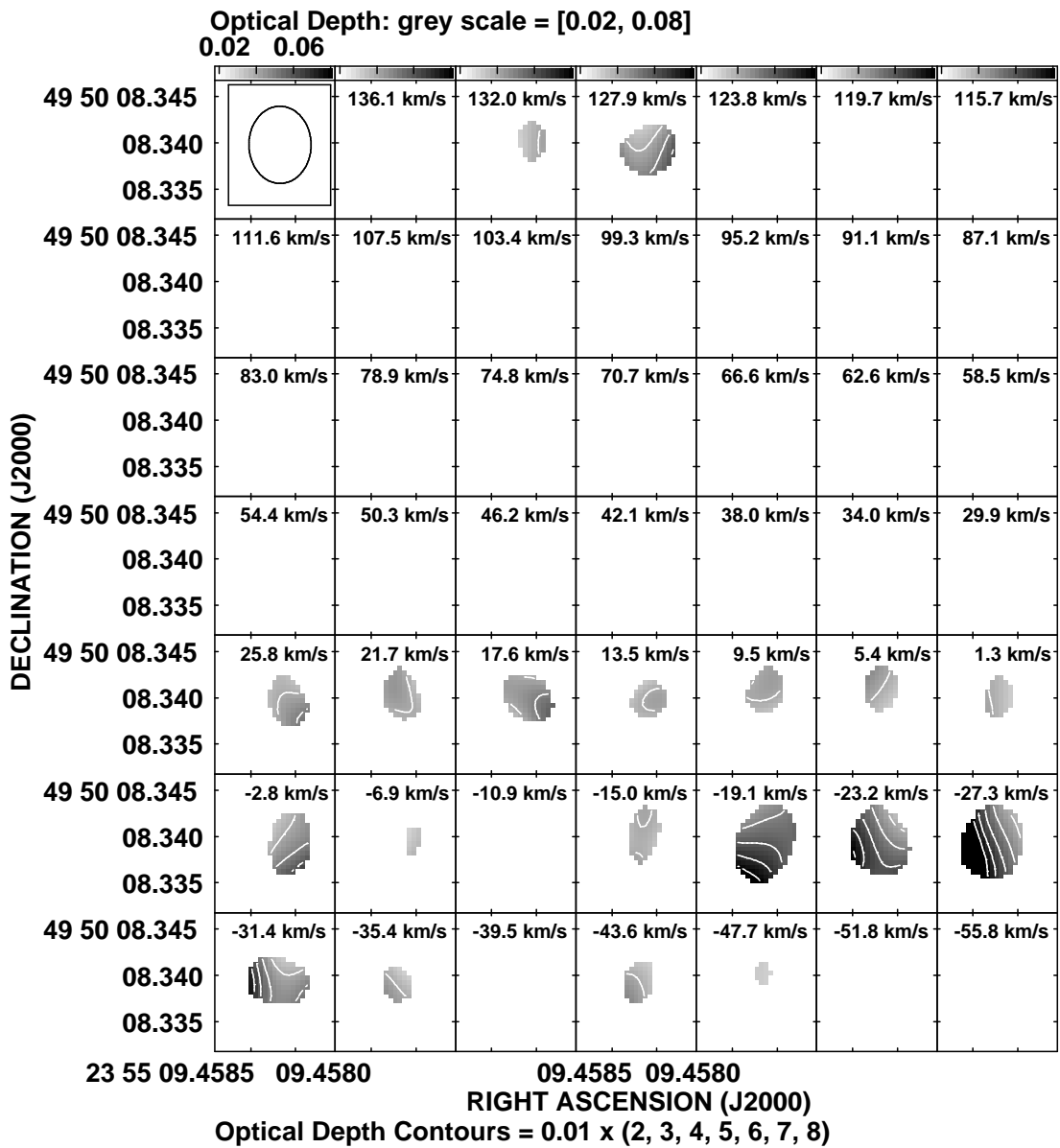} 
\vspace*{18.1cm}\caption{Optical depth channel maps 
(rest frame velocity is given). 
The area in each channel map corresponds to the 
central region of B2352+495 where \HI~was detected (compare the synthesized
beam size shown in the top-left panel with the synthesized
beam shown in Figure~1, bottom panel).
Only \HI~absorption with absolute magnitude greater than 4$\sigma$
was used to calculate the optical depth.}
\label{f2}
\end{figure}

\clearpage

\begin{figure}
\includegraphics{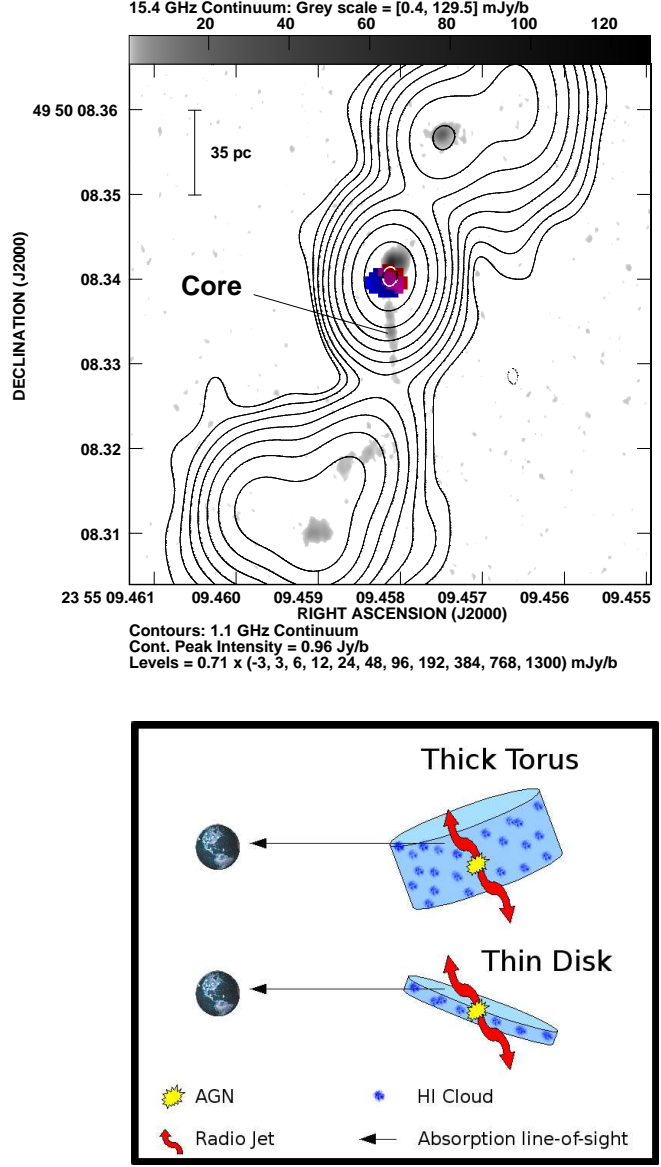} 
\vspace*{15.5cm}\caption{{\it Top:} Zoomed in image of the radio 
continuum in B2352+495 at 15.4$\,$GHz
(grey scale; Taylor et al.$\,$1996) and 1.14$\,$GHz (contours; this work).
Taylor et al.$\,$(1996) found that the center of activity (radio core) 
is located $\sim 20\,$pc south of the brightest 15.4$\,$GHz 
radio continuum source. In red and blue we show the \HI~optical depth 
channels that correspond to the rest frame velocities
of 17.6 and $-$31.4\kms~(see Figure~2; the opacity range of the red and 
blue channels is as in Figure~2, i.e., 0.02 to 0.08). 
We found a tentative \HI~velocity gradient that is almost 
perpendicular to the jet direction.
{\it Bottom:} Two possible
interpretations of the \HI~absorption in B2352+495: a thick torus
or a thin (possibly warped) disk. 
The red arrows represent the flow direction in the jets. The line
of sight to Earth (shown with the symbol taken from the VLBA/NRAO logo)
is in the plane of the picture. Radio continuum from parts of the jets
directed more towards Earth will appear brightened due to Doppler
boosting.}
\label{f3}
\end{figure}

\clearpage

\begin{figure}
\includegraphics{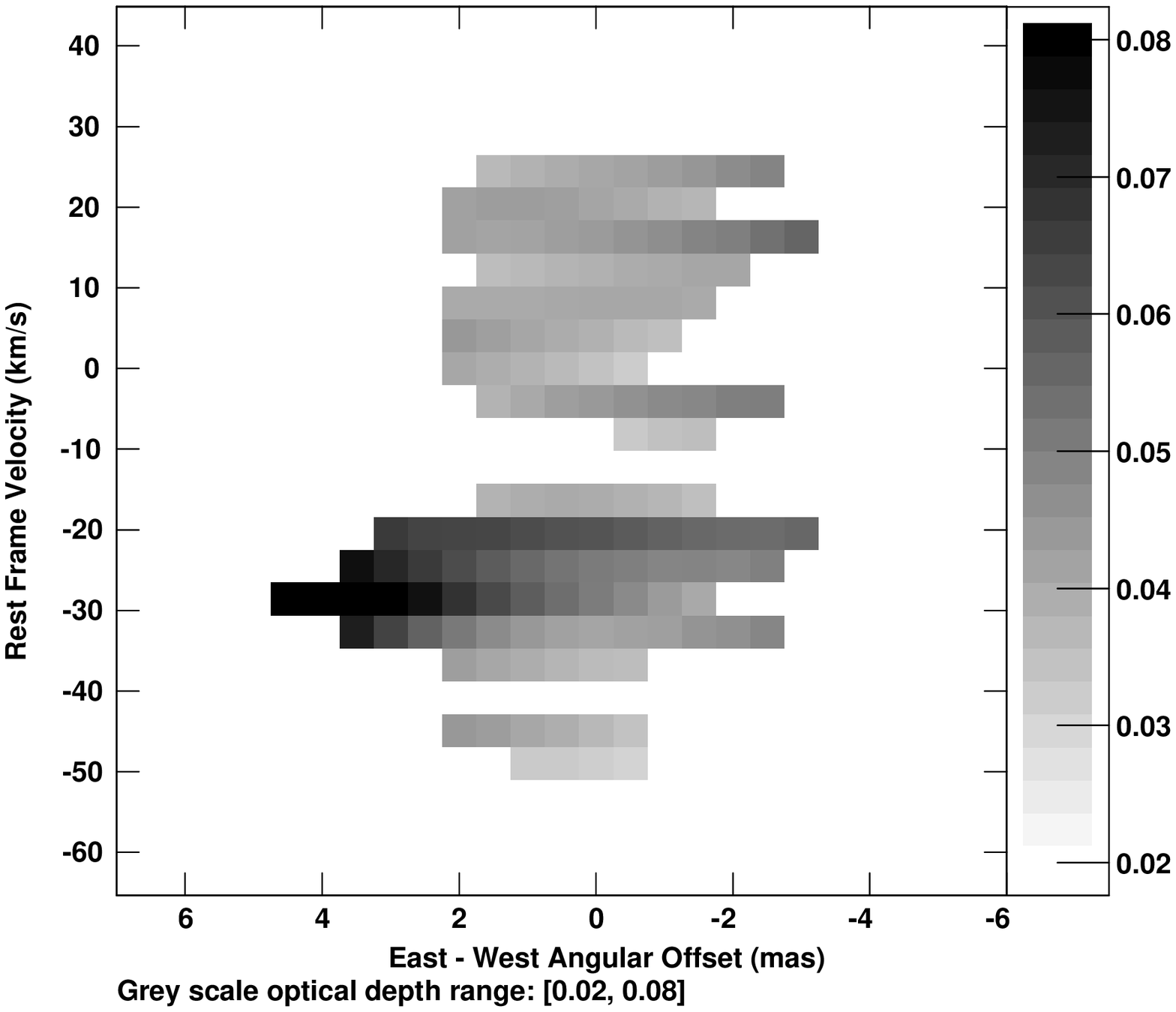} 
\vspace*{16.8cm}\caption{Position-velocity (P-V) diagram of the 
broad \HI~absorption component. The P-V diagram is along a 
east-west position angle passing through the center of the \HI~
absorption (Figure~1). Absorption along the declination axis was
averaged to increase signal-to-noise.}
\label{f4}
\end{figure}

\end{document}